\documentclass[floatfix,prb,twocolumn,showpacs,preprintnumbers,amsmath,
amssymb ] {revtex4}
\usepackage{graphicx}

\def\EPS{80mm}

\begin{document}

\title{ Thermophysical properties of warm dense hydrogen }
\author{ Bastian Holst and Ronald Redmer }
\address{ Universit\"at Rostock, Institut f\"ur Physik, 
          D-18051 Rostock, Germany }

\author{ Michael P. Desjarlais }
\affiliation{Pulsed Power Sciences Center, Sandia National Laboratories,
  Albuquerque, New Mexico 87185-1186, USA}

\date{\today}

\begin{abstract}
  We study the thermophysical properties of warm dense hydrogen using 
  quantum molecular dynamics simulations. New results are presented 
  for the pair distribution functions, the equation of state, 
  the Hugoniot curve, and the reflectivity. 
  We compare with available 
  experimental data and predictions of the chemical picture. Especially, 
  we discuss the nonmetal-to-metal transition which occurs at about 40 GPa
  in the dense fluid. 
\end{abstract}
\pacs{31.15.Ar, 61.20.Ja, 62.50.+p, 64.30.+t, 72.20.-i, 71.30.+h} 

\vspace*{5mm}

\maketitle

\section{Introduction}

Hydrogen is an essential element for models of stellar and planetary 
interiors~\cite{GUI99,GUI99b}. The isotopes deuterium and tritium are
considered as target materials (D-T gas) in inertial confinement fusion 
experiments~\cite{Lindl}. Therefore, numerous efforts have been made both 
experimentally and theoretically to understand the behavior of hydrogen, 
deuterium, and tritium in a wide range of densities and temperatures. 
In particular, progress in shock-wave experimental technique has allowed 
the systematic probing of the megabar pressure range, so that a sound database 
has been assembled within the last decade. Single or multiple shock-wave 
experiments have been performed for hydrogen (or deuterium) by using, e.g., 
high explosives~\cite{RUS-HE}, gas guns~\cite{NEL06}, 
pulsed power~\cite{KNU+01,KNU+03,KNU+04}, or high-power lasers~\cite{Nova}. 
The combination of high pressures and temperatures of several eV defines 
{\it warm dense matter}, a strongly correlated state relevant for planetary 
interiors which is characterized by partial ionization 
where the bound states exhibit a highly transient nature.

Furthermore, the enormous progress in computer capacity has allowed the
development and application of  
{\it ab initio} simulation techniques for warm dense matter such as Path 
Integral Monte Carlo (PIMC)~\cite{PIMC-Rev} or Quantum Molecular Dynamics 
(QMD) simulations~\cite{COL+00} which treat quantum effects and correlations 
systematically. These techniques give already highly predictive results for
a variety of problems and systems; see Ref.~\onlinecite{MATS05} for QMD
simulations. 

The equation of state (EOS) and derived quantities such as the Hugoniot 
curve, the sound velocity, or the Gr\"uneisen parameter are important 
material properties in this context. Furthermore, optical properties as, 
e.g., the reflectivity are closely related to the dielectric function 
which also determines the dc electrical conductivity in the static limit. 
All these quantities are used to characterize the unique behavior of warm 
dense hydrogen, especially for high pressures at, or exceeding, one megabar, 
where a transition from a nonconducting, molecular fluid to a mono-atomic
fluid with metallic-like conductivity occurs. Describing the disordered fluid 
in terms of solid state parameters, the fundamental band gap between the 
valence and conduction band decreases with the pressure and, subsequently, 
the electrical conductivity shows an exponential increase as is typical for 
thermally activated transport in semiconductors~\cite{NEL+92}. For 
pressures above 1.4~Mbar, conductivities of about 2000~$\Omega^{-1}$
cm$^{-1}$, as is characteristic for simple metallic fluids such as Cs, have
been observed experimentally around 3000 K~\cite{WEI+96,WEI+96b}, and band gap
closure has been claimed to be responsible for this nonmetal-to-metal
transition.

On the other hand, concepts of plasma physics have been applied to warm 
dense matter states~\cite{RED00}. For instance, the chemical picture gives 
a rather simple description by identifying stable bound states out of 
elementary particles as new composite particles. Hydrogen at normal 
conditions in this context is a molecular fluid. Free electrons are 
generated at high pressure by dissociation of molecules, 
H$_2 \rightleftharpoons$ 2 H, and a subsequent ionization of atoms, 
H $\rightleftharpoons$ e + p. This model yields already the strong 
increase of the conductivity with the pressure (pressure ionization). 
In addition, bound states contribute to conduction via hopping 
processes~\cite{RED+01}. The conceptual problem of all chemical models 
is the clear definition of bound states, the derivation of effective 
potentials between all species, and the calculation of cross sections 
for the respective scattering processes in a strongly correlated medium. 

QMD simulations are a powerful tool to describe warm dense 
matter~\cite{COL+95,LEN+97,COL+01,MPD+02,MPD03}. The combination of 
classical molecular dynamics for the ions and density functional theory 
(DFT) for the electrons allows one to consider correlation and quantum 
effects. Alternatively, wave packet simulations have been developed in 
which the electrons are represented on a semi-quantal level by wave 
packets (WPMD)~\cite{KLA+94,NAG+98,KNA+99,KNA+01,KNA+01b,JAK+07}. 

In this paper, we apply QMD simulations and calculate 
a broad spectrum of thermophysical properties of warm dense hydrogen. 
We determine EOS data for a wide region of densities and temperatures 
and compare with chemical models. We calculate the principal Hugoniot curve 
for liquid targets. The Kubo-Greenwood formula serves as a 
starting point for the evaluation of the dynamic conductivity 
$\sigma(\omega)$ from which the dielectric function 
$\varepsilon(\omega)$ and the reflectivity 
can be extracted. In addition, the electronic structure calculation within 
DFT yields the charge density distribution in the simulation box at every 
time step, and the molecular dynamics run gives valuable structural 
information via the ion-ion pair correlation function. This is important
for the identification and characterization of phase transitions such as
solid-liquid or liquid-plasma as well as for the nonmetal-to-metal 
transition.

\section{QMD simulations}

Within QMD simulations we perform molecular dynamics simulations with a quantum
mechanical treatment of the electrons by using density functional theory (DFT). 
This is based upon the theorems of Hohenberg and Kohn~\cite{Hohenberg} and
gives 
the electron density that minimizes the ground state energy of the system. 
It has been proven that this density is a unique functional of the effective 
potential $V_\text{eff}$.

From this formalism Kohn and Sham~\cite{Kohn} derived a computational scheme
which solves the problem for a fictious system of non-interacting particles
that leads to the same electron density. This scheme consists basically of 
solving the Kohn-Sham equations

\begin{eqnarray}
& \left[-\frac{\hbar^2}{2m}\nabla^2+V_{\text{eff}}(r)\right]
  \varphi_k(r) = \epsilon_k\varphi_k(r) ,&\\
& V_{\text{eff}}[\varrho(\mathbf{r})] = 
  \int\frac{\varrho(\mathbf{r'})e^2}{|\mathbf{r}-\mathbf{r'}|}d\mathbf{r'}
  - \sum_{k=1}^N \frac{Z_ke^2}{|\mathbf{r}-\mathbf{R}_k|}
  + V_{XC}[\varrho(\mathbf{r})] .\nonumber&
\label{eq:ks}
\end{eqnarray}

Our \textit{ab initio} quantum molecular dynamics simulations were performed
within Mermin's finite temperature density functional theory
(FT-DFT)~\cite{Mermin65}, which is implemented in the plane wave density functional 
code VASP (Vienna Ab Initio Simulation Package)~\cite{VASP,VASP2,VASP3}. 
We used the projector augmented wave potentials~\cite{PAW}
and did a generalized gradient approximation (GGA) using the parameterization 
of PBE~\cite{PBE}. Extensive test calculations, as performed already by
Desjarlais~\cite{MPD03}, have shown that the EOS data are 
dependent on the plane wave cutoff. A convergence of better than 1\% is secured for 
$E_\text{cut}=1200$~eV which was used in all actual calculations. The electronic 
structure calculations were performed for a given array of ion positions which are 
subsequently varied by the forces obtained within the DFT calculations via the 
Hellmann-Feynman theorem for each molecular dynamics step. This schema is repeated 
until the EOS measures are converged and a thermodynamic equilibrium is reached.

The simulations were done for 64~atoms in a supercell with periodic boundary
conditions. The temperature of the ions was controled by a Nos\'{e}
thermostat~\cite{Nose83} and the temperature of the electrons was fixed by Fermi
weighting the occupation of bands~\cite{VASP2}. The Brillouin zone was sampled
by evaluating the results at Baldereschi's mean value
point~\cite{Baldereschi73} which showed best agreement with a sampling of the
Brillouin zone using a higher number of \textbf{k}-points. The density of the
system was fixed by the size of the simulated supercell. To achieve a small
statistical error due to fluctuations the system was simulated 1000-1500 steps
further after reaching the thermodynamic equilibrium. The EOS data and pair
correlation functions were then obtained by averaging over all particles and 
simulation steps in equilibrium.
Similar calculations were performed recently for the thermophysical 
properties of warm dense helium~\cite{He-PRL} in order to verify the 
nonmetal-to-metal transition at high pressures.

\begin{figure}[htb]
\centering\includegraphics[width=\EPS]{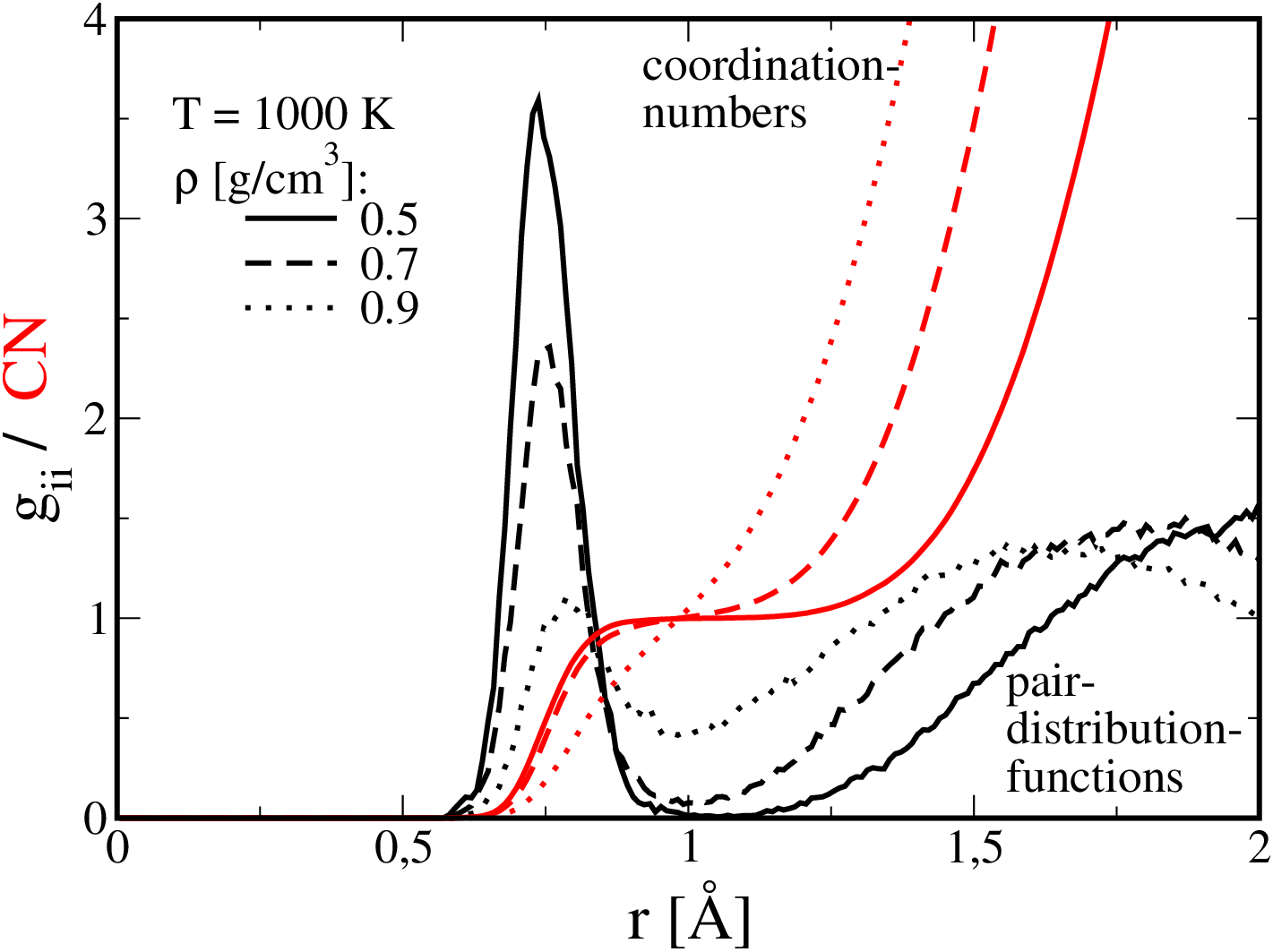}
    \caption{Proton-proton pair distribution function and corresponding
    coordination numbers according to Eq.~(\ref{eq:KN}) for 1000~K 
    and three densities.
\label{fig:PDF} }
\end{figure}

\begin{figure}[htb]
\centering\includegraphics[width=\EPS]{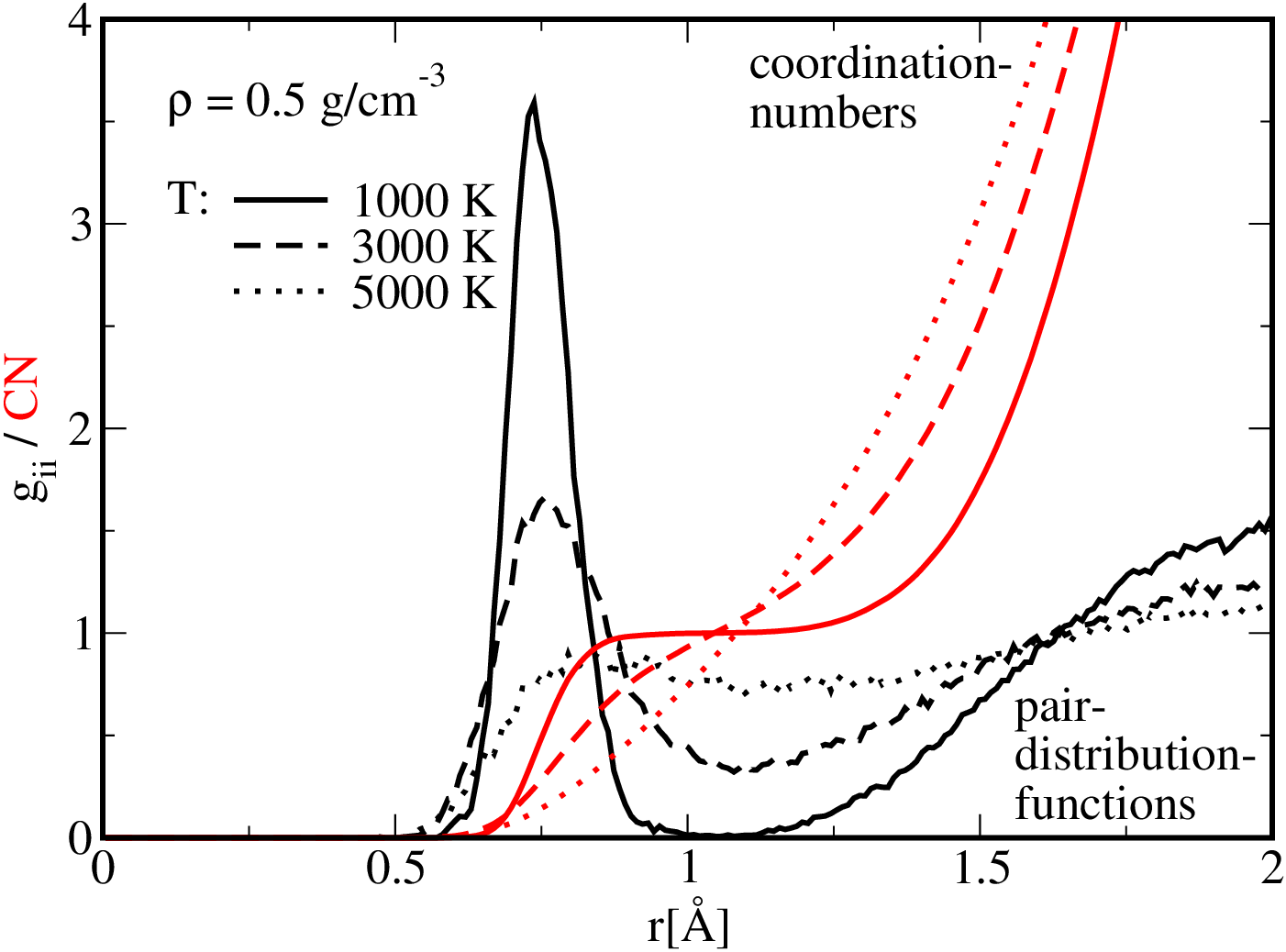}
    \caption{Proton-proton pair distribution function and corresponding
    coordination numbers according to Eq.~(\ref{eq:KN}) for 0.5~g/cm$^3$ 
    and three temperatures.
\label{fig:PDFtemp} }
\end{figure}

The zero-point vibrational energy of the H$_2$ molecules is not included in DFT 
calculations. In previous calculations, the energy $\frac{1}{2}h \nu_{vib}$ per 
molecule is simply added which is very important, especially at low temperatures 
and for the calculation of an exact initial internal energy for the reference 
state of the Hugoniot curve, which  is 0.0855~g/cm$^3$ at 20~K. To account for 
this quantum effect more sensitively for arbitrary temperatures, the
fraction 
of molecules has to be derived, e.g., for all states along the Hugoniot curve. 
This can be done via the coordination number 
\begin{equation} \label{eq:KN}
K(r)=\frac{N-1}{V}\int_0^r 4\pi r'^2 g(r')dr' ,
\end{equation}
which is a weighted integral over the pair correlation function $g(r)$ of the
ions. $N$ denotes the number of ions and $V$ the volume of the supercell in the 
simulation. The doubled value of $K$ at the maximum of the molecular peak in 
$g(r)$, which is found around $r=0.748$~\AA{}, is then equal to the fraction of 
ions bound to a molecule and twice the amount of molecules in the supercell. 
An example is shown in Fig.~\ref{fig:PDF} where the increasing dissociation
with higher density can be seen. In Fig.~\ref{fig:PDFtemp} we show the thermal
dissociation; the molecular peak dissappears with increasing
temperature at constant density. Note that the peak is thermally broadened.

The dissociation degree is calculated for a number of isotherms and then 
approximated by a Fermi function which has two adjustable parameters. These 
parameters can be represented by temperature-dependent functions so that the 
dissociation degree and, subsequently, the contribution of molecules to the 
zero-point internal energy are determined for arbitrary temperatures. The 
results show that molecules can be neglected above 10,000~K.

\begin{figure}[htb]
\centering\includegraphics[width=\EPS]{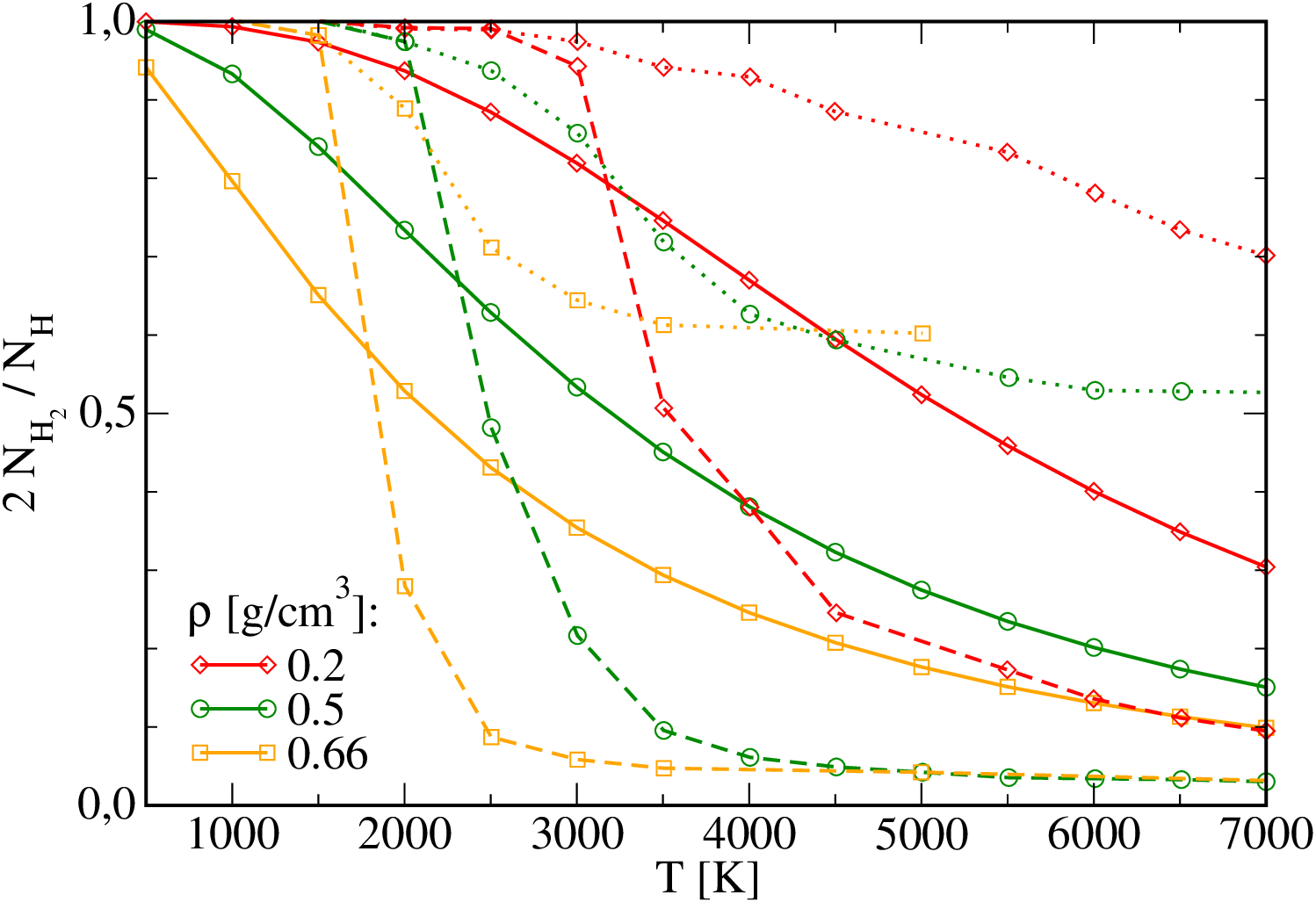}
\caption{Ratio of hydrogen molecules with respect to the total number of 
protons for three densities: Our coordination number method (solid) 
is compared with the pair-counting method of 
Vorberger~\textit{et al.}~\cite{VOR+07} (dotted). Their result counting 
only pairs with a lifetime longer than ten H$_2$ vibrational periods is 
also given (dashed line).
\label{fig:Hmolratio} }
\end{figure}

We compare the resulting dissociation degree with that derived by Vorberger 
\textit{et al.}~\cite{VOR+07} in Fig.~\ref{fig:Hmolratio}. They counted all
pairs of atoms in a range of 1.8~a$_B$ as atoms. In a second step they reduced
the number of molecules by counting only those pairs that are stable for longer 
than ten vibrational periods. In all three cases the amount of molecules is 
lower for higher densities und the molecules disappear at higher temperatures 
due to thermal dissociation. This picture shows that the dissociation degree 
depends strongly on the definition of the term {\it molecule} in the 
warm dense matter region. Our alternative method gives a
smoother behavior of the dissociation degree which starts at lower 
temperatures and is in between the two cases described by 
Vorberger~\textit{et al.}~\cite{VOR+07} at higher temperatures.

\section{Results for the EOS and Hugoniot curves}

\begin{figure}[htb]
\centering\includegraphics[width=\EPS]{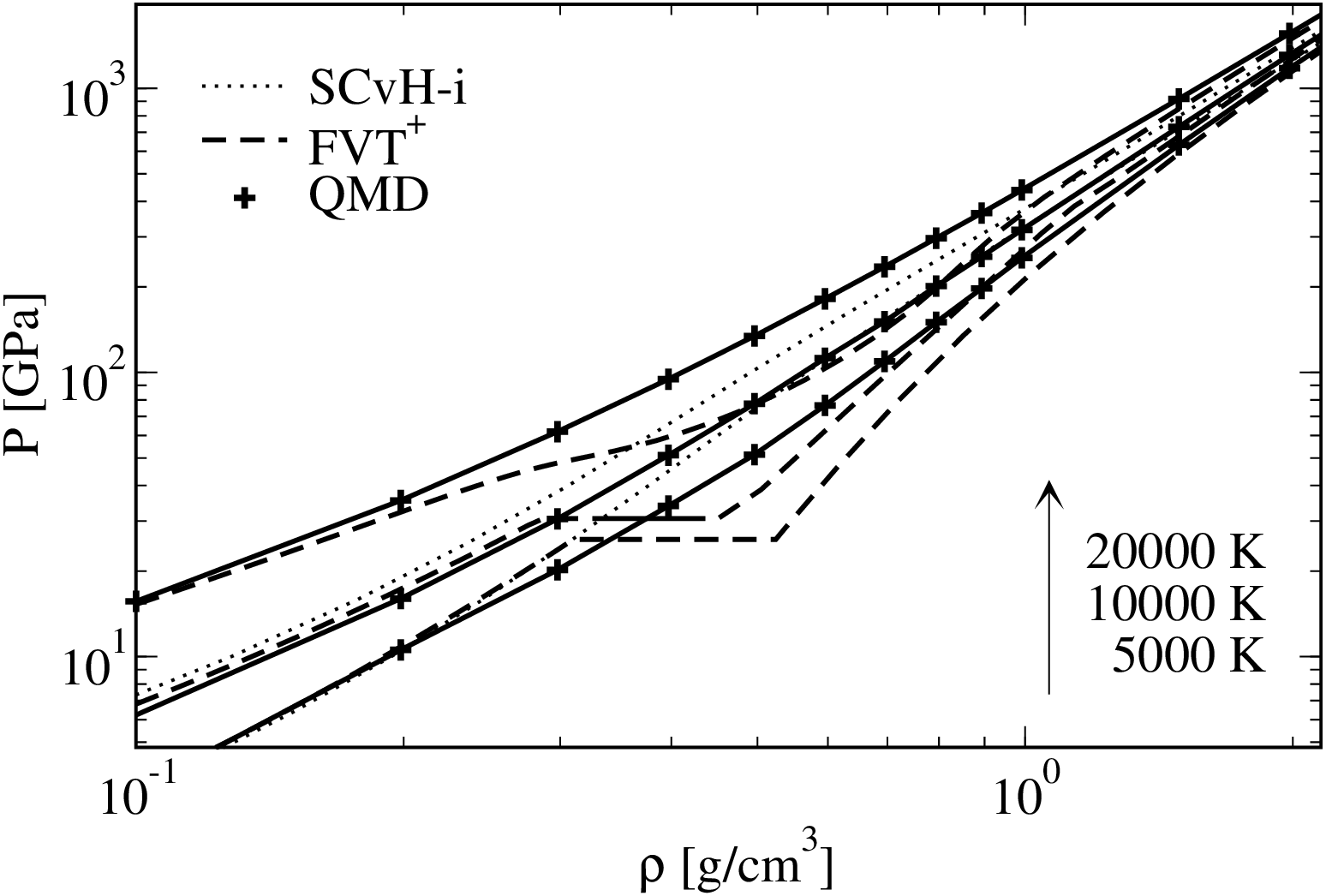}
\caption{Thermal EOS for warm dense hydrogen (pressure isotherms):
QMD data are compared with the chemical models FVT$^+$~\protect\cite{Holst+07}
and SCvH-i~\protect\cite{SCVH}. 
\label{fig:EOS-PVT} }
\end{figure}

We show the thermal EOS of warm dense hydrogen in Fig.~\ref{fig:EOS-PVT}. 
The isotherms of the pressure show a systematic behavior in terms of the density 
and temperature. We find no indication of a plasma phase transition (PPT) which
would result in an instability of the EOS isotherms that would need to be
treated by a Maxwell
construction~\cite{Beule01}. The absence of a PPT is in contrast to results of 
chemical models which use, e.g., Fluid Variational Theory 
(FVT)~\cite{FVT1,FVT2,Holst+07} or liquid state perturbation theory~\cite{SCVH}. 
Chemical models are based on a free energy minimization schema for a mixture of 
hydrogen atoms, molecules, and a plasma in chemical equilibrium. Correlations are 
taken into account based on effective two-particle potentials. The description 
of the free charged particles (plasma) is done beyond the Debye-H\"uckel 
approximation by using efficient Pad\'{e} formulas which are valid for a wide 
region of densities and temperatures. 

In Fig.~\ref{fig:EOS-PVT} our QMD results are compared with the chemical models 
FVT~\cite{Holst+07} and SCvH-i~\cite{SCVH}. The EOS derived by Saumon 
{\it et al.}~\cite{SCVH} shows also a PPT (SCvH-ppt data set). The modified SCvH-i 
data set shown here avoids the PPT by using an interpolation through the 
instability region. Therefore, both data sets can be used to study the influence 
of a PPT on interior models of giant planets such as Jupiter.
Consistent chemical models yield the correct low-temperature and low-density limit 
and agree with our QMD results there. A good agreement is also found in the 
high-density limit where a nearly temperature independent behavior
characteristic 
of a degenerate plasma is found. At medium densities the pressure isotherms of 
FVT and SCvH-i lie well below the QMD data; the deviations amount up to 25\%.

We have encountered a region with $(\partial P/\partial T)_V<0$, which was
previously reported by Vorberger \textit{et al.}~\cite{VOR+07}. It can be
related to the rapid dissociation transition at low temperatures.

Following Lenosky \textit{et al.}~\cite{LEN+97a} and Beule \textit{et
al.}~\cite{BEU+99} we fit smooth functions for the
pressure $P$ and the internal energy $U$ as an expansion in terms of density
$\rho$ and temperature $T$ to the given results of the QMD simulations.
The pressure is split into an ideal and an interaction contribution:
\begin{equation}
 P=P^{id}+P^{int}=\frac{\rho k_B T}{m_H}+P^{int}(\rho,T) .
\end{equation}
The QMD data for the pressure $P$ given in kbar can be interpolated by the
following expansion for the interaction contribution:
\begin{equation}
P^{int}(\rho,T)=\left(A_1(T)+A_2(T)\rho\right)^{A_0(T)} ,\label{eq:Pexp}
\end{equation}
\begin{equation}
A_i(T) = a_{i0} 
\exp\left(-\left(\frac{T-a_{i1}}{a_{i2}}\right)^2\right)+a_{i3}+a_{i4}T .
\label{eq:Pcoeff}
\end{equation}
The coefficients a$_{ik}$ are summarized in Tab.~\ref{tab:Pcoeff}.

\begin{table}
\caption{\label{tab:Pcoeff}Coefficients $a_{ik}$ in the expansion for the
pressure $P^{int}$ according to Eqs.~(\ref{eq:Pexp}) and (\ref{eq:Pcoeff}).}
\begin{ruledtabular}
\begin{tabular}{cccccc}
i&$a_{i0}$&$a_{i1}$&$a_{i2}$&$a_{i3}$&$a_{i4}$\\
\hline
0&0.2234&2919.84&3546.67&1.94023&1.11316$\cdot10^{-6}$\\
1&14.7586&2117.98&4559.17&-17.9538&4.88041$\cdot10^{-4}$\\
2&-33.8469&2693.63&4159.13&70.582&-2.8848$\cdot10^{-4}$
\end{tabular}
\end{ruledtabular}
\end{table}

The QMD data for the specific internal energy $u=U/m$ given in kJ/g can be
interpolated by a similar expansion:
\begin{equation}
u=\sum_{j=0}^4B_j(T)\rho^j ,\label{eq:Uexp}
\end{equation}
\begin{equation}
B_j(T) = b_{j0} 
\exp\left(-\left(\frac{T-b_{j1}}{b_{j2}}\right)^2\right)+b_{j3}+b_{j4}T .
\label{eq:Ucoeff}
\end{equation}
The expansion coefficients $b_{jk}$ are given in Tab.~\ref{tab:Ucoeff}.

\begin{table}
\caption{\label{tab:Ucoeff}Coefficients $b_{jk}$ in the expansion for the
specific internal energy $u$ according to Eqs.~(\ref{eq:Uexp}) and
(\ref{eq:Ucoeff}).}
\begin{ruledtabular}
\begin{tabular}{cccccc}
j&$b_{j0}$&$b_{j1}$&$b_{j2}$&$b_{j3}$&$b_{j4}$\\
\hline
0&-33.8377&2154.38&3696.89&-300.446&1.77956$\cdot10^{-2}$\\
1&55.8794&3174.39&2571.21&56.222&-3.56234$\cdot10^{-3}$\\
2&-30.0376&3174.02&2794.39&87.3659&2.0819$\cdot10^{-3}$\\
3&5.57328&3215.51&2377.23&-13.1622&-3.84004$\cdot10^{-4}$\\
4&-0.3236&3245.48&2991.45&0.682152&2.19862$\cdot10^{-5}$\\
\end{tabular}
\end{ruledtabular}
\end{table}

The expansions (\ref{eq:Pexp}) and (\ref{eq:Uexp}) reproduce the 
{\it ab initio} QMD data within 5\% accuracy in a density range from 
0.5~g/cm$^3$ to 5~g/cm$^3$ between 500~K and 20000~K and can easily 
be applied in planetary models or hydrodynamic simulations for warm 
dense matter. The expansions fulfill thermodynamic consistency 
expressed by the relation 
\begin{equation}
 P-T\left(\frac{\partial P}{\partial T}\right)_V
=-\left(\frac{\partial U}{\partial V}\right)_T
\end{equation}
within 15\% accuracy which is mainly due to the deviations from 
the QMD data itself.

\begin{figure}[htb]
\centering\includegraphics[width=\EPS]{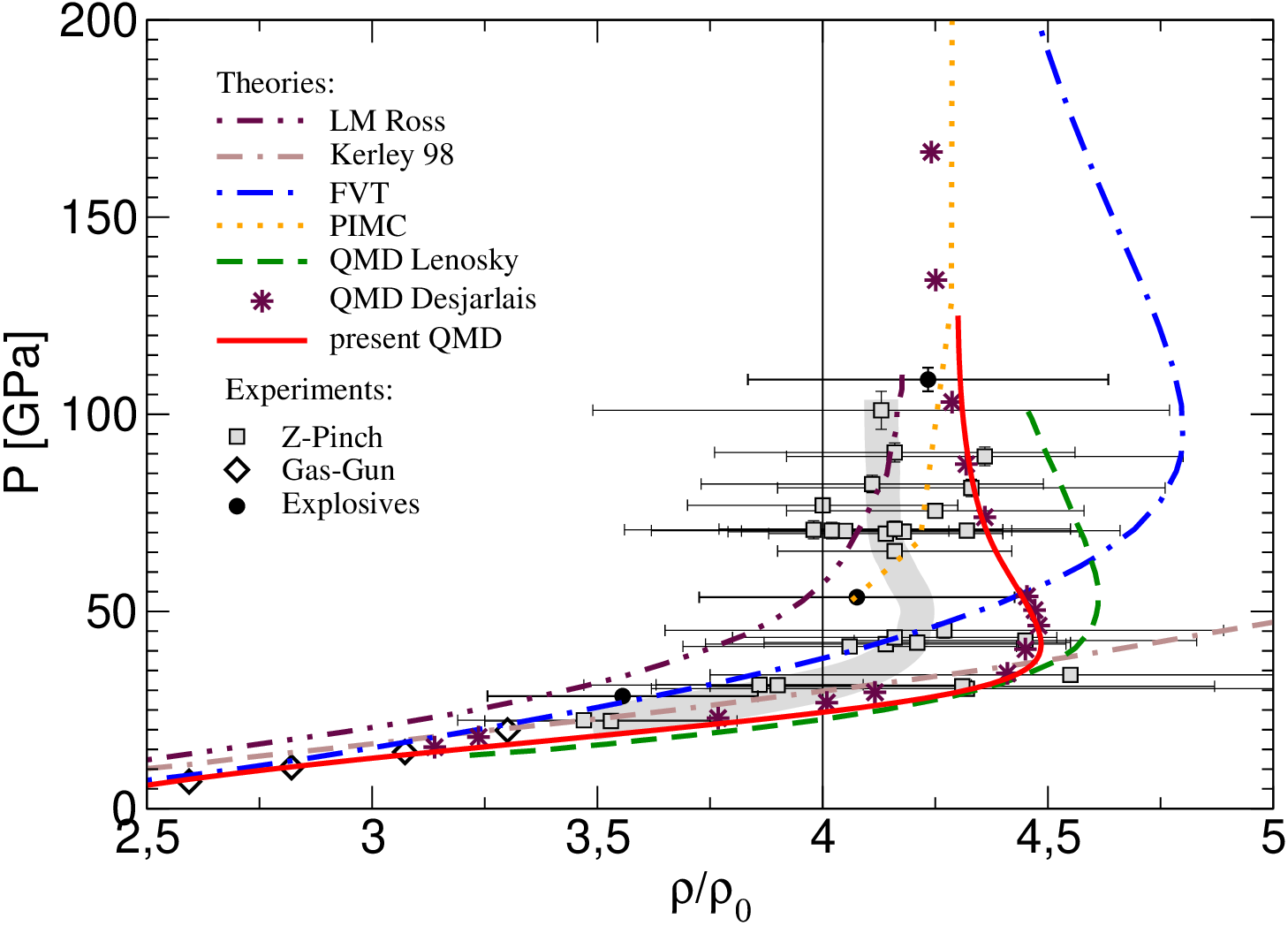}
    \caption{ 
Principal Hugoniot curve for hydrogen. The results of this work (solid line)
are compared with previous QMD results of 
Lenosky {\it et al.}~\protect\cite{Lenosky+00} (dashed) and 
Desjarlais~\protect\cite{MPD03} (stars),  
PIMC simulations~\protect\cite{MIL+00} (dotted), 
the linear mixing model of Ross\cite{ROS98}
(dot-dash-dashed), the model of Kerley~\protect\cite{KER03} (dot-dot-dashed)
and the chemical model FVT~\protect\cite{FVT1} (dot-dashed). 
Experiments: Gas gun~\protect\cite{NEL+83} (diamonds), 
Sandia Z~machine~\protect\cite{KNU+04} (grey squares; grey line: running average 
through the $u_s$-$u_p$ data), high explosives~\protect\cite{Boriskov+05} (black circles). 
\label{fig:Hugo-princip} 
}
\end{figure}

A crucial measure for theoretical EOS data is the principal Hugoniot curve
which is plotted in Fig.~\ref{fig:Hugo-princip}. It describes all possible
final states ($\rho,P,u$) of shock wave experiments according to the Hugoniot
equation
\begin{equation}\label{eq:hugo}
u-u_0=\frac{1}{2}(P+P_0)(\frac{1}{\rho_0}-\frac{1}{\rho_0}) 
\end{equation}
starting at the same initial conditions ($\rho_0,P_0,u_0$). For the hydrogen
principal Hugoniot curve, the initial density is $\rho_0=0.0855$~g/cm$^3$ and
the initial internal energy $u_0=-314$~kJ/g at a temperature of 20~K. The
initial pressure $P_0$ can be neglected because of the high pressure of the
final state.

Shock wave experiments have been performed for deuterium using gas guns~\cite{NEL+83}, 
magnetically launched flyer plates at Sandia's Z~machine~\cite{KNU+04} or high 
explosives (HE)~\cite{Boriskov+05}. These experiments indicate a maximum compression 
of 4.25 at about 50~GPa. 

Another series of laser-driven experiments~\cite{Nova} shows systematic deviations 
from the experiments quoted above. Especially, a maximum compression of 6 has been 
reported at about 1~Mbar. According to the unanimous evaluation of the shock-wave
experimental data for molecular liquids~\cite{NEL06}, we compare our QMD data in 
Fig.~\ref{fig:Hugo-princip} only with the data sets mentioned above.

The systematic increase of the cutoff energy $E_\text{cut}$ in QMD simulations 
from 500~eV~\cite{Lenosky+00} to 1200~eV\cite{MPD03}
has lead to fully converged results in agreement with the experimental points.
The consideration of the zero-point vibrations of the H$_2$ molecules along the
entire Hugoniot curve yields a very good agreement of QMD data with the gas gun 
experiments~\cite{NEL+83} expecially for low pressures. The calculated Hugoniot
curve has a 
maximum compression of 4.5 which is slightly higher than the HE and Z
experiments indicate (about 4.25). This is an agreement of about 5\% accuracy
which can be translated into an accuracy of about 1\% in the measured shock
and particle velocity, which is in the range of the systematic errors in
the experiments.
The compression decreases with higher pressures and
temperatures and 
reaches the correct high-temperature limit as given by the PIMC 
simulations~\cite{MIL+02}. The QMD curve lies slightly below the experimental data 
for compression rates between 3 and 4 which could be due to the known band gap 
problem of DFT in GGA.
The FVT curve~\cite{FVT1} is shown as a representative of chemical models which,
in general, show a higher compressibility well beyond 4.5.

Also shown is the linear mixing result of Ross~\cite{ROS98}. This curve shows  a
sixfold compression and is not in agreement with the shown experiments.  
The curve of Kerley~\cite{KER03} has a maximum compression of 4.25, like the
experiments indicate, but the pressure  is there
slightly higher than the results of the QMD simulations.

\section{Dynamic conductivity, reflectivity and dc conductivity}

The dynamic conductivity $\sigma(\omega)$ is derived from the Kubo-Greenwood
formula:~\cite{Kubo,Greenwood}
\begin{eqnarray}\label{eq:kg}
\sigma(\omega) &=& \frac{2\pi e^2\hbar^2}{3 m^2 \omega \Omega}
\sum_{\mathbf k} W({\mathbf k}) \sum_{j=1}^N\sum_{i=1}^N\sum_{\alpha=1}^3 
\left[ F(\epsilon_{i,{\mathbf k}})-F(\epsilon_{j,{\mathbf k}})\right] \nonumber\\
&& \times |\langle\Psi_{j,{\mathbf k}}|\nabla_\alpha|\Psi_{i,{\mathbf k}}\rangle|^2
\delta(\epsilon_{j,{\mathbf k}}-\epsilon_{i,{\mathbf k}}-\hbar\omega) ,
\end{eqnarray}
where $e$ is the electron charge and $m$ its mass. The summations over $i$
and $j$ run over $N$ descrete bands considered in the electronic structure
calculation for the cubic supercell volume $\Omega$. The three spatial
directions are averaged by the $\alpha$ sum.
$F(\epsilon_{i,{\mathbf k}})$ describes the occupation of the $i$th band
corresponding to the energy $\epsilon_{i,{\mathbf k}}$ and the wavefunction
$\Psi_{i,{\mathbf k}}$ at ${\bf k}$. The $\delta$-function has to be broadened
because a discrete energy spectrum results from the finite simulation
volume~\cite{MPD+02}. Integration over the Brillouin zone is performed by
sampling special ${\mathbf k}$ points~\cite{Monkhorst},
where $W({\mathbf k})$ is the respective weighting factor.
We used Baldereschi's  mean value
point~\cite{Baldereschi73} to reach a convergence of better than 10\%
accuracy.

Optical properties can be derived from the frequency-dependent conductivity 
Eq.~(\ref{eq:kg}). The standard method is to obtain the imaginary part via the 
Kramers-Kronig relation
\begin{equation}
\sigma_2(\omega)=-\frac{2}{\pi} \text{P}\int
\frac{\sigma_1(\nu)\omega}{(\nu^2-\omega^2)}d\nu ,
\label{eq:kkr}
\end{equation}
$\text{P}$ is the prinicipal value of the integral. The dielectric function 
can be calculated directly with the conductivity:
\begin{align}\label{eq:df}
\epsilon_1(\omega)=1-\frac{1}{\epsilon_0 \omega}\sigma_2(\omega),\\
\epsilon_2(\omega)=\frac{1}{\epsilon_0 \omega}\sigma_1(\omega).
\end{align}

The square of the index of refraction contains the real part $n$ and the 
imaginary part $k$ is equal to the dielectric function which leads to the 
following relations:
\begin{align}\label{eq:ri}
n(\omega)=\frac{1}{2}\sqrt{|\epsilon(\omega)|+|\epsilon_1(\omega)|},\\
k(\omega)=\frac{1}{2}\sqrt{|\epsilon(\omega)|-|\epsilon_1(\omega)|}.
\end{align}

The index of refraction is then used to calculate optical propersties such as 
the reflectivity $r$:
\begin{equation}\label{eq:reflex}
r(\omega)=\frac{[1-n(\omega)]^2+k(\omega)^2}{[1+n(\omega)]^2+k(\omega)^2}.
\end{equation}

\begin{figure}[htb]
\centering\includegraphics[width=\EPS]{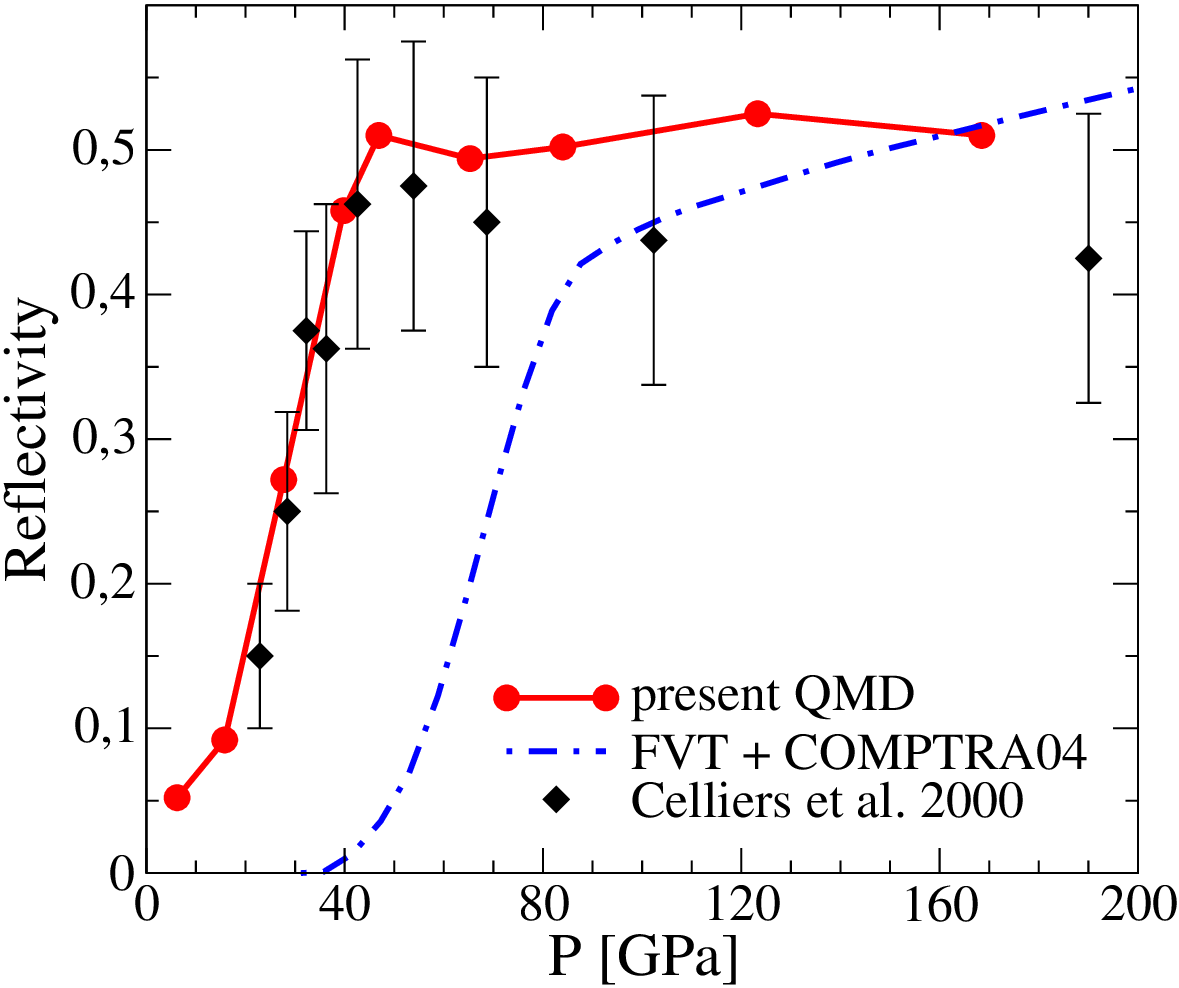}
    \caption{ Reflectivity for a wavelength of 808 nm along the Hugoniot curve
of hydrogen: 
    QMD results are compared with experimental data of Celliers 
    {\it et al.}~\protect\cite{Celliers} and predictions of the 
    chemical model FVT~\protect\cite{APS} using the COMPTRA 
    code~\protect\cite{Comptra04}. 
\label{fig:reflex} }
\end{figure}

We compare our {\it ab initio} results with reflectivities measured along the 
Hugoniot curve~\cite{Celliers} in Fig.~\ref{fig:reflex}; the agreement is 
excellent. The change of the hydrogen reflectivity with the pressure can be 
interpreted as a gradual transition from a molecular insulating fluid through 
an atomic fluid above 20~GPa where the atoms have strongly fluctuating bonds 
with next neighbors~\cite{COL+01} to a dense, almost fully ionized plasma 
with a reflectivity of about 50-60~\% at high pressures above 40~GPa. 
The chemical model~\cite{APS} shows also this qualitative behavior but the 
abrupt increase of the reflectivity occurs at a higher density. This shows 
the difficulties of the chemical models in finding the correct shifts of the 
dissociation and ionization energies as function of density and temperature 
and, thus, the location of the nonmetal-to-metal transition. However, the 
limits of a molecular fluid at low pressures and of a fully ionized plasma 
at high pressures are incorporated in a reasonable way.

\section{Conclusion}

We have performed {\it ab initio} QMD simulations to study thermophysical
properties of hydrogen under extreme conditions. 
As a result we obtained highly converged EOS data which are relevant for
modeling giant planets and for the understanding of the fundamental behavior of
hydrogen at high pressure. The deviations between our QMD data and chemical 
models amount up to 25\%. We have constructed smooth fit functions 
for the QMD data for the pressure and the internal energy which can be used 
easily in, e.g., hydrodynamic simulations for warm dense hydrogen and in 
astrophysical applications.

The results show a smooth transition from a molecular liquid to an atomic fluid
of metal-like state. There were no signs of a PPT which is predicted by
other models. 

With these EOS results we have calculated the principal Hugoniot curve which
is in agreement with dynamic experiments and has the correct high-temperature limit
as given by PIMC simulations. 

We obtained optical properties using the Kubo-Greenwood formula.
The reflectivity along the Hugoniot curve is in excellent agreement with
experiments. The results show the occurrence of a nonmetal-to-metal
transition at about 40 GPa.

\begin{acknowledgements}

We thank P.M.~Celliers, W.~Ebeling, V.E.~Fortov, M.~French, A.~H\"oll, 
A.~Kietzmann, W.D.~Kraeft, T.R.~Mattsson, B.~Militzer, V.B.~Mintsev, 
N.~Nettelmann, H.~Reinholz, G.~R\"opke, N.A.~Tahir, V.Ya.~Ternovoi, 
C.~Toepffer, and G.~Zwicknagel
for stimulating discussions and for providing us with their data. 
This work was supported by the Deutsche Forschungsgemeinschaft within
the SFB~652 and the grant mvp00006 of the High Performance Computing 
Center North (HLRN). We acknowledge support of the computer center of 
the University of Rostock (URZ). 
Sandia is a multiprogram laboratory operated by Sandia Corporation, a
Lockheed Martin Company, for the United States Department of Energy's
National Nuclear Security Administration under contract DE-AC04-94AL85000.

\end{acknowledgements}


\bibliography{qmd3}

\end{document}